\documentclass[11pt]{article}

\input tcilatex
\begin{document}

\title{\textbf{RICCI COLLINEATIONS FOR NON-DEGENERATE, DIAGONAL AND SPHERICALLY
SYMMETRIC RICCI TENSORS.}}
\author{G. Contreras\thanks{%
In memoriam}, \\
\textit{Laboratorio de F\'{\i }sica Te\'{o}rica,}\\
\textit{Departamento de F\'{\i }sica, Facultad de Ciencias, }\\
\textit{Universidad de Los Andes, M\'{e}rida 5101, VENEZUELA} \and L.A.
N\'{u}\~{n}ez\thanks{\texttt{nunez@ciens.ula.ve}}, \\
\textit{Centro de Astrof\'{\i }sica Te\'{o}rica,}\\
\textit{Departamento de F\'{\i }sica, Facultad de Ciencias,}\\
\textit{Universidad de Los Andes, M\'{e}rida 5101, VENEZUELA }\\
and \textit{Centro Nacional de C\'{a}lculo Cient\'{\i }fico}\\
\textit{Universidad de Los Andes }\textsc{(CeCalCULA)}\textit{,}\\
\textit{Corporaci\'{o}n Parque Tecnol\'{o}gico de M\'{e}rida, }\\
\textit{M\'{e}rida 5101, VENEZUELA} \and U. Percoco\thanks{\texttt{%
upercoco@ciens.ula.ve}} \\
\textit{Centro de Astrof\'{\i }sica Te\'{o}rica,}\\
\textit{Departamento de F\'{\i }sica, Facultad de Ciencias,}\\
\textit{Universidad de Los Andes, M\'{e}rida 5101, VENEZUELA.}}
\maketitle

\begin{abstract}
The expression of the vector field generator of a Ricci Collineation for
diagonal, spherically symmetric and non-degenerate Ricci tensors is
obtained. The resulting expressions show that the time and radial first
derivatives of the components of the Ricci tensor can be used to classify
the collineation, leading to 64 families.\newline
Some examples illustrate how to obtain the collineation vector.
\end{abstract}

\section{Introduction.}

General Relativity provides a rich arena to use symmetries in order to
simplify and understand the natural relation between geometry and matter
furnished by the Einstein equations. Symmetries of geometrical/physical
relevant quantities of this theory are known as \textit{Collineations} and,
in general, can be represented as: $\pounds _{\vec{\xi}}\phi =F$, where $%
\phi $ and $F$ are two geometric objects, $\vec{\xi}$ is the vector field
generating the symmetry and $\pounds _{\vec{\xi}}$ the Lie Derivative along
the congruence generated by $\vec{\xi}$. The hierarchy and relations among
these symmetries are presented in the seminal work of Katzin and Levine \cite
{Katzin69} (see Fig.1 below). The particular position occupied by the Ricci
Collineation (RC), defined by $(\pounds _{\xi }R)_{ab}=0$ , at the top of
the hierarchical tree, and its close relation with the Energy-Momentum
tensor, strongly motivate the study of \textit{Proper and Improper Ricci
Collineations}. Collineations can be \textit{proper} or \textit{improper}. A
collineation of a given type is said to be\textit{\ proper} if it does not
belong to any of the subtypes. Clearly, in solving for example equation $%
(\pounds _{\xi }R)_{ab}=0$, solutions representing improper collineations
can be found. However, in order to be related to a particular conservation
law, and its corresponding constant of the motion, the ``properness'' of the
collineation type must be assured.

Ricci Collineations for static spherically symmetric space-times have been
studied recently by various authors \cite{Paqui93} \cite{Paqui94} \cite
{Paqui95} \cite{Paqui96} \cite{Bertol}. This work deals with the RC vector
for \textit{dynamic} (i.e. non-static) spherically symmetric space-times. A
detailed analysis of the $(\pounds _{\xi }R)_{ab}=0$ suggests a
classification of RC based on the vanishing the time and radial first
derivatives of the components of the Ricci tensor. This approach, as can be
seen in the next section, leads to 64 families of RC, each one distinguished
by a set of vanishing first derivative of the components of the Ricci
tensor; these results are summarized in Table 1.

Section three gives some examples of RC for some of the families listed in
Table 1, with special emphasis on three specific FRW type metric tensors.%
\newline

\newpage

\begin{picture}(400,400)(0,55)
\put(-5,40){\makebox(400,400)}

\put(155,430){\framebox(50,20){\small CRC}}
\put(155,430){\makebox(50,20) [lt]{\small 17}}
\put(0,390){\framebox(40,20){\small WPC}}
\put(40,400){\vector(3,-1){120}}
\put(0,390){\makebox(40,20) [lt]{\small 1}}
\put(160,390){\framebox(40,20){\small RC}}
\put(160,390){\makebox(40,20) [lt]{\small 4}}
\put(180,410){\vector(0,1){20}}
\put(320,390){\framebox(70,20){\small WConfC}}
\put(320,400){\vector(-3,-1){120}}
\put(320,390){\makebox(70,20) [lt]{\small 12}}

\put(160,350){\framebox(40,20){\small CC}}
\put(160,350){\makebox(40,20) [lt]{5}}
\put(180,370){\vector(0,1){20}}
\put(30,355){\line(1,0){130}}
\put(30,355){\vector(0,1){35}}
\put(200,355){\vector(3,1){120}}
\put(200,355){\vector(1,-4){51}}
\put(0,310){\framebox(40,20){\small PC}}
\put(0,310){\makebox(40,20) [lt]{2}}
\put(20,330){\vector(0,1){60}}
\put(40,320){\vector(1,-1){30}}
\put(100,310){\framebox(60,20){\small SCC}}
\put(100,310){\makebox(60,20) [lt]{6}}
\put(160,320){\line(1,0){10}}
\put(170,320){\vector(0,1){30}}
\put(320,310){\framebox(70,20){\small Conf C}}
\put(320,310){\makebox(70,20) [lt]{13}}
\put(320,320){\vector(-2,-1){59}}

\put(50,270){\framebox(50,20){\small SPC}}
\put(50,270){\makebox(50,20) [lt]{3}}
\put(50,280){\line(-1,0){30}}
\put(20,280){\vector(0,1){30}}
\put(75,290){\line(0,1){30}}
\put(75,320){\vector(1,0){25}}
\put(120,270){\framebox(50,20){\small NC}}
\put(120,270){\makebox(50,20) [lt]{15}}
\put(150,270){\vector(1,-1){20}}
\put(230,270){\framebox(80,20){\small S Conf C}}
\put(230,270){\makebox(80,20) [lt]{10}}
\put(310,280){\line(1,0){30}}
\put(340,280){\vector(0,1){30}}
\put(160,230){\framebox(60,20){\small SNC}}
\put(160,230){\makebox(60,20) [lt]{16}}
\put(180,250){\vector(0,1){100}}
\put(160,240){\line(-1,0){30}}
\put(130,240){\vector(0,1){30}}

\put(160,170){\framebox(50,20){\small AC}}
\put(160,170){\makebox(50,20) [lt]{7}}
\put(180,190){\vector(0,1){40}}
\put(160,180){\line(-1,0){85}}
\put(75,180){\vector(0,1){90}}
\put(210,180){\line(1,0){40}}
\put(250,180){\vector(0,1){90}}
\put(320,170){\framebox(70,20){\small Conf M}}
\put(320,170){\makebox(70,20) [lt]{14}}
\put(355,190){\vector(0,1){120}}

\put(230,130){\framebox(80,20){\small S Conf M}}
\put(230,130){\makebox(80,20) [lt]{11}}  
\put(270,150){\vector(0,1){120}}
\put(310,140){\line(1,0){45}}
\put(355,140){\vector(0,1){30}}

\put(160,80){\framebox(50,20){\small HM}}
\put(160,80){\makebox(50,20) [lt]{8}}
\put(185,100){\vector(0,1){70}}
\put(210,90){\line(1,0){60}}
\put(270,90){\vector(0,1){40}}

\put(160,40){\framebox(50,20){\small M}}
\put(160,40){\makebox(50,20) [lt]{9}}
\put(185,60){\vector(0,1){20}}
\end{picture}
{\tiny {.}}\vspace{0.1in}\vspace{0.1in}

\begin{center}
{\footnotesize {\textbf{Fig.1.}\, Relations between symmetries.\\[0pt]
The leant arrows relate symmetries for which: $R_{ij}=0$.}}
\end{center}

{\footnotesize {\noindent{{1. WPC - Weyl Proyective Collineation: $\pounds
W_{jkl}^i=0$ \thinspace $(n>2)$.}}\newline
2. PC - Proyective Collineation: $\pounds \Gamma _{jk}^i=\delta _j^i\phi
_{;k}+\delta _k^i\phi _{;j}$.\newline
3. SPC - Special Proyective Collineation: $\pounds \Gamma _{jk}^i=\delta
_j^i\phi _{;k}+\delta _k^i\phi _{;j}$,\thinspace $\phi _{;jk}=0$.\newline
4. RC - Ricci Collineation: $\pounds R_{ij}=0$.\newline
5. CC - Curvature Collineation: $\pounds R_{jkl}^i=0$.\newline
6. SCC - Special Curvature Collineation: $(\pounds \Gamma _{jk}^i)_{;l}=0$.%
\newline
7. AC - Affine Collineation: $\pounds \Gamma _{jk}^i=0$.\newline
8. HM - Homothetic Motion: $\pounds g_{ij}=2\sigma g_{ij}$,\thinspace $%
\sigma =ctte.$.\newline
9. M - Motion:$\pounds g_{ij}=0$.\newline
10. S Conf C - Special Conformal Collineation:$\pounds \Gamma _{jk}^i=\delta
_j^i\sigma _{;k}+\delta _k^i\sigma _{;j}-g_{jk}g^{il}\sigma _{;l}$, $\sigma
_{;jk}=0$.\newline
11. S Conf M - Special Conformal Motion: $\pounds g_{ij}=2\sigma g_{ij}$,
\thinspace $\sigma _{;jk}=0$.\newline
12. W Conf C - Weyl Conformal Collineation: $\pounds $ $C_{jkl}^i=0$,
\thinspace $(n>3)$.\newline
13. Conf C - Conformal Collineation: $\pounds \Gamma _{jk}^i=\delta
_j^i\sigma _{;k}+\delta _k^i\sigma _{;j}-g_{jk}g^{il}\sigma _{;l}$.\newline
14. Conf M - Conformal Motion:$\pounds g_{ij}=2\sigma g_{ij}$.\newline
15. NC - null geodesic Collineation: $\pounds \Gamma
_{jk}^i=g_{jk}g^{im}\psi _{;m}$.\newline
16. SNC - Special null geodesic Collineation:$\pounds \Gamma
_{jk}^i=g_{jk}g^{im}\psi _{;m}$, \thinspace $\psi _{;jk}=0$.\newline
17. CRC - Contracted Ricci Collineation:\thinspace $g^{ij}\pounds R_{ij}=0$}}%
\newline

\section{The expression of the RC vector field.}

Recently, the following result has been obtained by J. Carot et al. \cite
{Carot97}: \textit{Therefore, the proper RC of a spherically symmetric
space-time whose Ricci tensor is non-degenerate, are of the form:} 
\begin{equation}
\vec{\xi}=\xi ^{t}(t,r)\partial _{t}+\xi ^{r}(t,r)\partial _{r}.
\end{equation}
Thus, we will integrate the system of equations $(\pounds _{\vec{\xi}%
}R)_{ab}=0$ for a RC vector field with just two components ( $\xi ^{t}$ and $%
\xi ^{r}$ ) depending only of $t$ and $r$. Consideration of this
collineation vector $\vec{\xi}$ does not preclude that another class of
collineations, other than proper RC, could be obtained. Indeed: the form of
the most general RC vector is the one given above plus linear combinations,
with constant coefficients, of the Killing Vectors for spherical symmetry 
\cite{Carot97}.

We consider a diagonal Ricci Tensor $R_{ab}$ written in those coordinates
where (\cite{Kramer} pag.163 ): 
\begin{equation}
ds^{2}=-e^{2\nu (t,r)}dt^{2}+e^{2\lambda (t,r)}dr^{2}+Y^{2}(t,r)(d\theta
^{2}+sen^{2}\theta d\phi ^{2}).
\end{equation}
Then, the system of equations:
\begin{equation}
C_{ab}=(\pounds _{\vec{\xi}}R)_{ab}=0,\qquad \mathrm{with\quad }%
\,a,b=t,r,\theta ,\phi   \label{RicciCol}
\end{equation}
reduces to: 
\begin{eqnarray}
C_{tt} &=&\xi ^{t}\partial _{t}A+\xi ^{r}\partial _{r}A+2A\xi _{\ ,t}^{t}=0
\label{RicciColtt} \\
C_{tr} &=&A\xi _{\ ,r}^{t}+B\xi _{\ ,t}^{r}=0 \\
C_{rr} &=&\xi ^{t}\partial _{t}B+\xi ^{r}\partial _{r}B+2B\xi _{\ ,r}^{r}=0
\\
C_{\theta \theta } &=&\xi ^{t}\partial _{t}C+\xi ^{r}\partial _{r}C=0
\label{RicciColthth}
\end{eqnarray}
where the following notation is used: $A=R_{tt}$, $B=R_{rr}$, and $%
C=R_{\theta \theta }$.

From the above equations ((\ref{RicciColtt}) through (\ref{RicciColthth}))
we can get,
\begin{eqnarray}
0 &=&BC_{tt}+AC_{rr}  \nonumber \\
&=&B\xi ^{t}\partial _{t}A+B\xi ^{r}\partial _{r}A+2AB\partial _{t}\xi
^{t}+A\xi ^{t}\partial _{t}B+A\xi ^{r}\partial _{r}B+2AB\partial _{r}\xi ^{r}
\nonumber \\
&=&(B\partial _{t}A+A\partial _{t}B)\xi ^{t}+(B\partial _{r}A+A\partial
_{r}B)\xi ^{r}+2AB(\partial _{t}\xi ^{t}+\partial _{r}\xi ^{r})
\label{Comb3}
\end{eqnarray}
Setting: 
\begin{equation}
\xi ^{a}=(AB)^{-1/2}\eta ^{a}
\end{equation}
with $A,B\neq 0$ \thinspace and \thinspace $\eta =\eta (t,r)$ we have, 
\begin{equation}
\partial _{a}\xi ^{a}=-1/2(AB)^{-3/2}(B\partial _{a}A+A\partial _{a}B)\eta
^{a}+(AB)^{-1/2}\partial _{a}\eta ^{a}  \label{derxi}
\end{equation}

Now from equations.(\ref{Comb3}) and (\ref{derxi}) it is clear that: 
\begin{equation}
2(AB)^{1/2}\partial _{a}\eta ^{a}=0
\end{equation}
with the following solution: 
\begin{equation}
\eta ^{a}=\epsilon ^{ab}\partial _{b}\phi ,\hspace{0.2in}\phi =\phi (t,r)
\end{equation}
where,\thinspace  $\epsilon ^{tr}=1$ , \thinspace $\epsilon ^{rt}=-1$ and
\thinspace  $\epsilon ^{tt}$= $\epsilon ^{rr}=0$

Let us now consider $C_{tr}$: 
\begin{equation}
C_{tr}=A\partial _{r}[\eta ^{t}(AB)^{-1/2}]+B\partial _{t}[\eta
^{r}(AB)^{-1/2}]=0
\end{equation}
Differentiation with respect to $\phi $ and multiplying the result by $%
2(AB)^{3/2}$ we obtain: 
\begin{equation}
C_{tr}=A\left( -\frac{\partial _{r}(AB)}{AB})\partial _{r}\phi +2\partial
_{rr}\phi \right) +B\left( \frac{\partial _{t}(AB)}{AB}\partial _{t}\phi
-2\partial _{tt}\phi \right) =0
\end{equation}
From $C_{tt}$, $C_{\theta \theta }$ and $C_{tr}$ we have now the following
system of equations: 
\begin{eqnarray}
A\left( -\frac{\partial _{r}(AB)}{AB}\partial _{r}\phi +2\partial _{rr}\phi
\right) +B\left( \frac{\partial _{t}(AB)}{AB}\partial _{t}\phi -2\partial
_{tt}\phi \right)  &=&0  \label{RicciColNew1} \\
-\frac{\partial _{t}B}{B}\partial _{r}\phi -\frac{\partial _{r}A}{A}\partial
_{t}\phi +2\partial _{rt}\phi  &=&0  \label{RicciColNew2} \\
\partial _{t}C\partial _{r}\phi -\partial _{r}C\partial _{t}\phi  &=&0
\label{RicciColNew3}
\end{eqnarray}
and the form of the RC emerges as: 
\begin{equation}
\xi ^{t}=\frac{\partial _{r}\phi }{\sqrt{AB}},\hspace{0.2in}\mathrm{and}%
\quad \xi ^{r}=-\frac{\partial _{t}\phi }{\sqrt{AB}}\ .  \label{Colvect}
\end{equation}

From equations. (\ref{RicciColNew1}), (\ref{RicciColNew2}) and (\ref
{RicciColNew3}) we see that the partial derivatives of the components of the
Ricci tensor will appear in the expression of $\vec{\xi}$. In order to
classify all this sort of RC vectors, the afore mentioned equations suggest
(almost obligate) to consider the vanishing of one or more derivatives of
the components of the Ricci tensor as classifying parameters. With this
criterion we get the 64 cases shown in TABLE 1.

\section{Calculating a RC.}

In order to illustrate how to extract information from TABLE 1, we give some
detailed examples.

\subsection{{\protect\underline{Family 1: $\partial _{r}A=0$ .}}}

The equations (\ref{RicciColNew1}), (\ref{RicciColNew2}) and (\ref
{RicciColNew3}) take the following expression in this case:
\begin{eqnarray}
\partial _{r}\phi \frac{\partial _{t}C}{\partial _{r}C}-\partial _{t}\phi 
&=&0  \label{Fam1E1} \\
-\frac{\partial _{t}B}{B}\partial _{r}\phi +2\partial _{tr}\phi  &=&0
\label{Fam1E2} \\
A\left( -\frac{\partial _{r}(B)}{B}\partial _{r}\phi +2\partial _{rr}\phi
\right) +B\left( \frac{\partial _{t}(AB)}{AB}\partial _{t}\phi -2\partial
_{tt}\phi \right)  &=&0\ .  \label{Fam1E3}
\end{eqnarray}
From equations (\ref{Fam1E2}) and (\ref{Fam1E3}) we get: 
\begin{equation}
\partial _{r}\phi =\sqrt{B}f(r),\hspace{0.2in}\mathrm{and\quad }\partial
_{t}\phi =\frac{\partial _{t}C}{\partial _{r}C}\sqrt{B}f(r)\ .
\end{equation}
Because $\partial _{rt}\phi =\partial _{rt}\phi ,$ we have 
\begin{equation}
f(r)=e^{\Delta ^{(1)}}\ ,  \label{Fam1fr}
\end{equation}
where 
\begin{equation}
\Delta ^{(1)}=\frac{1}{2}\int \frac{B}{A}\frac{\partial _{t}C}{\partial _{r}C%
}\left( -\frac{\partial _{t}(A)}{A}+2\frac{\partial _{t}\left( \frac{%
\partial _{t}C}{\partial _{r}C}\right) }{\frac{\partial _{t}C}{\partial _{r}C%
}}\right) {d}r\ .
\end{equation}
Finally considering equations (\ref{Colvect}), the RC vector has the
following components: 
\begin{equation}
_{(1)}\xi ^{t}=\frac{e^{\Delta ^{(1)}}}{\sqrt{A}},\hspace{0.2in}\mathrm{%
and\quad }_{(1)}\xi ^{r}=-\frac{\partial _{t}C}{\partial _{r}C}\frac{%
e^{\Delta ^{(1)}}}{\sqrt{A}}  \label{ColvectF1}
\end{equation}

\subsection{\protect\underline{\textbf{{Family 7:\thinspace $\partial _{r}A=0
$\thinspace and \thinspace $\partial _{r}B=0$.}}}}

Equations (\ref{RicciColNew1}), (\ref{RicciColNew2}) and (\ref{RicciColNew3}%
) take the following expression in this case:
\begin{eqnarray}
\partial _{r}\phi \frac{\partial _{t}C}{\partial _{r}C}-\partial _{t}\phi 
&=&0  \label{Fam7E1} \\
-\frac{\partial _{t}B}{B}\partial _{r}\phi +2\partial _{tr}\phi  &=&0
\label{Fam7E2} \\
2A\partial _{rr}\phi +B\left( \frac{\partial _{t}(AB)}{AB}\partial _{t}\phi
-2\partial _{tt}\phi \right)  &=&0  \label{Fam7E3}
\end{eqnarray}
\begin{tabular}{|c|c||c|c|}
\hline
Family & vanishing derivatives & Family & vanishing derivatives \\ \hline
1 & $\partial _{r}A=0$ & 33 & $\partial _{t}A$, $\partial _{t}B$, $\partial
_{r}B=0$ \\ \hline
2 & $\partial _{r}B=0$ & 34 & $\partial _{t}A$, $\partial _{t}B$, $\partial
_{r}C=0$ \\ \hline
3 & $\partial _{r}C=0$ & 35 & $\partial _{t}A$, $\partial _{t}B$, $\partial
_{t}C=0$ \\ \hline
4 & $\partial _{t}A=0$ & 36 & $\partial _{t}A$, $\partial _{t}C$, $\partial
_{r}A=0$ \\ \hline
5 & $\partial _{t}B=0$ & 37 & $\partial _{t}A$, $\partial _{t}C$, $\partial
_{r}B=0$ \\ \hline
6 & $\partial _{t}C=0$ & 38 & $\partial _{t}A$, $\partial _{t}C$, $\partial
_{r}C=0$ \\ \hline
7 & $\partial _{r}A$, $\partial _{r}B=0$ & 39 & $\partial _{r}A$, $\partial
_{t}B$, $\partial _{t}C=0$ \\ \hline
8 & $\partial _{r}A$, $\partial _{r}C=0$ & 40 & $\partial _{t}B$, $\partial
_{t}C$, $\partial _{r}B=0$ \\ \hline
9 & $\partial _{r}B$, $\partial _{r}C=0$ & 41 & $\partial _{t}B$, $\partial
_{t}C$, $\partial _{r}C=0$ \\ \hline
10 & $\partial _{t}A$, $\partial _{t}B=0$ & 42 & $\partial _{r}A$, $\partial
_{t}A$, $\partial _{r}B$, $\partial _{r}C=0$ \\ \hline
11 & $\partial _{t}A$, $\partial _{t}C=0$ & 43 & $\partial _{r}A$, $\partial
_{t}B$, $\partial _{r}B$, $\partial _{r}C=0$ \\ \hline
12 & $\partial _{t}B$, $\partial _{t}C=0$ & 44 & $\partial _{r}A$, $\partial
_{r}B$, $\partial _{t}C$, $\partial _{r}C=0$ \\ \hline
13 & $\partial _{t}A$, $\partial _{r}A=0$ & 45 & $\partial _{r}A$, $\partial
_{t}A$, $\partial _{r}B$, $\partial _{t}B=0$ \\ \hline
14 & $\partial _{r}A$, $\partial _{t}B=0$ & 46 & $\partial _{r}A$, $\partial
_{t}A$, $\partial _{r}B$, $\partial _{t}C=0$ \\ \hline
15 & $\partial _{r}A$, $\partial _{t}C=0$ & 47 & $\partial _{r}A$, $\partial
_{t}B$, $\partial _{r}B$, $\partial _{t}C=0$ \\ \hline
16 & $\partial _{t}A$, $\partial _{r}B=0$ & 48 & $\partial _{r}A$, $\partial
_{t}A$, $\partial _{t}B$, $\partial _{r}C=0$ \\ \hline
17 & $\partial _{t}A$, $\partial _{r}C=0$ & 49 & $\partial _{r}A$, $\partial
_{t}A$, $\partial _{t}C$, $\partial _{r}C=0$ \\ \hline
18 & $\partial _{t}B$, $\partial _{r}B=0$ & 50 & $\partial _{r}A$, $\partial
_{t}B$, $\partial _{t}C$, $\partial _{r}C=0$ \\ \hline
19 & $\partial _{t}B$, $\partial _{r}C=0$ & 51 & $\partial _{t}A$, $\partial
_{t}B$, $\partial _{r}B$, $\partial _{r}C=0$ \\ \hline
20 & $\partial _{t}C$, $\partial _{r}C=0$ & 52 & $\partial _{t}A$, $\partial
_{r}B$, $\partial _{t}C$, $\partial _{r}C=0$ \\ \hline
21 & $\partial _{r}B$, $\partial _{t}C=0$ & 53 & $\partial _{t}B$, $\partial
_{r}B$, $\partial _{t}C$, $\partial _{r}C=0$ \\ \hline
22 & $\partial _{r}A$, $\partial _{r}B$, $\partial _{r}C=0$ & 54 & $\partial
_{t}A$, $\partial _{r}A$, $\partial _{t}B$, $\partial _{t}C=0$ \\ \hline
23 & $\partial _{r}A$, $\partial _{r}B$, $\partial _{t}A=0$ & 55 & $\partial
_{t}A$, $\partial _{r}B$, $\partial _{t}B$, $\partial _{t}C=0$ \\ \hline
24 & $\partial _{r}A$, $\partial _{r}B$, $\partial _{t}B=0$ & 56 & $\partial
_{t}A$, $\partial _{r}C$, $\partial _{t}B$, $\partial _{t}C=0$ \\ \hline
25 & $\partial _{r}A$, $\partial _{r}B$, $\partial _{t}C=0$ & 57 & $\partial
_{t}A$, $\partial _{r}A$, $\partial _{t}B$, $\partial _{r}B$, $\partial
_{r}C=0$ \\ \hline
26 & $\partial _{r}A$, $\partial _{t}A$, $\partial _{r}C=0$ & 58 & $\partial
_{t}A$, $\partial _{r}A$, $\partial _{r}B$, $\partial _{t}C$, $\partial
_{r}C=0$ \\ \hline
27 & $\partial _{r}A$, $\partial _{t}B$, $\partial _{r}C=0$ & 59 & $\partial
_{r}A$, $\partial _{t}B$, $\partial _{r}B$, $\partial _{t}C$, $\partial
_{r}C=0$ \\ \hline
28 & $\partial _{r}A$, $\partial _{t}C$, $\partial _{r}C=0$ & 60 & $\partial
_{t}A$, $\partial _{r}A$, $\partial _{r}B$, $\partial _{t}C$, $\partial
_{r}B=0$ \\ \hline
29 & $\partial _{t}A$, $\partial _{r}B$, $\partial _{r}C=0$ & 61 & $\partial
_{t}A$, $\partial _{r}A$, $\partial _{t}B$, $\partial _{t}C$, $\partial
_{r}C=0$ \\ \hline
30 & $\partial _{r}B$, $\partial _{t}B$, $\partial _{r}C=0$ & 62 & $\partial
_{t}A$, $\partial _{t}B$, $\partial _{r}B$, $\partial _{t}C$, $\partial
_{r}C=0$ \\ \hline
31 & $\partial _{r}B$, $\partial _{t}C$, $\partial _{r}C=0$ & 63 & $\partial
_{t}A$, $\partial _{r}A$, $\partial _{r}B$, $\partial _{t}B$, $\partial _{t}C
$, $\partial _{r}C=0$ \\ \hline
32 & $\partial _{r}A$, $\partial _{t}A$, $\partial _{t}B=0$ & 64 & $\partial
_{t}A$, $\partial _{r}A$, $\partial _{r}B$, $\partial _{t}B$, $\partial _{t}C
$, $\partial _{r}C\neq 0$ \\ \hline
\end{tabular}

{\tiny {.}}\hspace{1.5in} \textbf{TABLE 1:} Families of RC vectors.

Again, from equations (\ref{Fam7E2}) and (\ref{Fam7E3}) we get: 
\begin{equation}
\partial _{r}\phi =\sqrt{B}f(r),\hspace{0.2in}\mathrm{and\quad }\partial
_{t}\phi =\frac{\partial _{t}C}{\partial _{r}C}\sqrt{B}f(r)
\end{equation}
and $\partial _{rt}\phi =\partial _{rt}\phi ,$ yields, 
\begin{equation}
\frac{\partial _{r}f(r)}{f(r)}=-\frac{B}{2}\left( \frac{\partial _{t}(AB)}{AB%
}\frac{\partial _{t}C}{\partial _{r}C}-2\partial _{t}\left( \frac{\partial
_{t}C}{\partial _{r}C}\right) -\frac{\partial _{t}C}{\partial _{r}C}\frac{%
\partial _{t}B}{B}\right) 
\end{equation}
so that, 
\begin{equation}
f(r)=e^{\Delta ^{(7)}}
\end{equation}
where 
\begin{equation}
\Delta ^{(7)}=\frac{B}{2}\int {\frac{\partial _{t}C}{\partial _{r}C}\left( -%
\frac{\partial _{t}(A)}{A}+2\frac{\partial _{t}\left( \frac{\partial _{t}C}{%
\partial _{r}C}\right) }{\frac{\partial _{t}C}{\partial _{r}C}}\right) dr}=%
\frac{\Delta ^{(1)}}{A}
\end{equation}
The components of the RC vector are: 
\begin{equation}
_{(7)}\xi ^{t}=\frac{e^{\Delta ^{(7)}}}{\sqrt{A}},\hspace{0.2in}\mathrm{%
and\quad }_{(7)}\xi ^{r}=-\frac{\partial _{t}C}{\partial _{r}C}\frac{%
e^{\Delta ^{(7)}}}{\sqrt{A}}  \label{ColvectF7}
\end{equation}

\subsection{{\protect\underline{\textbf{{Family 64:\thinspace  $\partial
_{c}R_{ab}\neq 0$\thinspace  for \thinspace  $c=t,r$.}}}}}

From equations (\ref{RicciColNew2}), (\ref{RicciColNew3}) and (\ref{Colvect}%
) we get, 
\begin{equation}
_{(64)}\xi ^{t}=\frac{e^{\Delta ^{(64)}}}{\sqrt{A}},\hspace{0.2in}\mathrm{%
and\quad }_{(64)}\xi ^{r}=\frac{-\partial _{t}C}{\partial _{r}C}\frac{%
e^{\Delta ^{(64)}}}{\sqrt{A}}
\end{equation}
where 
\begin{equation}
\Delta ^{(64)}=\frac{1}{2}\int \frac{\partial _{r}A}{A}\frac{\partial _{t}C}{%
\partial _{r}C}{d}t
\end{equation}
and the \textit{restriction equation} emerging from equation.(\ref
{RicciColNew1}): 
\begin{equation}
A\left( -\frac{\partial _{r}A}{A}+2\partial _{r}\Delta ^{(64)}\right) +B%
\frac{\partial _{t}C}{\partial _{r}C}\left( \frac{\partial _{t}A}{A}-2\frac{%
\partial _{t}\left( \frac{\partial _{t}C}{\partial _{r}C}\right) }{\frac{%
\partial _{t}C}{\partial _{r}C}}-\frac{\partial _{r}A}{A}\frac{\partial _{t}C%
}{\partial _{r}C}\right) =0
\end{equation}
with 
\begin{equation}
\partial _{r}C\neq 0
\end{equation}

\subsection{\textbf{Three examples of FRW type metric tensors.}}

\subsubsection{\textbf{First Example:}}

Consider the following line element \cite{NPV},
\begin{equation}
ds^{2}=dt^{2}-F^{2}(t)[\frac{1}{1-kr^{2}}dr^{2}+r^{2}d\theta
^{2}+r^{2}sen^{2}\theta d\phi ^{2}]
\end{equation}
L. A. N\'{u}\~{n}ez \textit{et al}. have supposed a collineation vector of
the following form:
\begin{equation}
\vec{\xi}=(\xi ^{t}(t,r),\xi ^{r}(t,r),0,0)
\end{equation}
The components of the Ricci tensor in this case are:\newline
\begin{equation}
A=-3F_{,00}/F,\hspace{0.1in}B=\Delta /(1-kr^{2}),\hspace{0.1in}C=r^{2}\Delta
,\hspace{0.1in}R_{33}=Csen^{2}\theta 
\end{equation}
where: $\Delta (t)=2k+2(F_{,0})^{2}+FF_{,00}$.

This metric tensor belongs to the family number \textbf{1} of the \textbf{%
TABLE 1} ($\partial _{r}A~=~0$). The components of the RC vector $\vec{\xi}$
are the following (see equations (\ref{ColvectF1})): 
\begin{equation}
_{(1)}\xi ^{t}=\frac{w(t)\sqrt{1-kr^{2}}}{\sqrt{A}}exp\left( -\frac{\partial
_{t}\Delta }{4kA}\left( -\frac{\partial _{t}A}{A}+2\frac{\partial _{t}\left( 
\frac{\partial _{t}\Delta }{\Delta }\right) }{\left( \frac{\partial
_{t}\Delta }{\Delta }\right) }\right) \right) 
\end{equation}
\begin{equation}
_{(1)}\xi ^{r}=-\frac{r\partial _{t}\Delta }{2\Delta }\frac{w(t)\sqrt{%
1-kr^{2}}}{\sqrt{A}}exp\left( -\frac{\partial _{t}\Delta }{4kA}\left( -\frac{%
\partial _{t}A}{A}+2\frac{\partial _{t}\left( \frac{\partial _{t}\Delta }{%
\Delta }\right) }{\left( \frac{\partial _{t}\Delta }{\Delta }\right) }%
\right) \right) 
\end{equation}
\newline

Taking into account the equation (\ref{Fam1fr}), we can assume that 
\begin{equation}
w(t)=const\qquad \mathrm{and\quad }\left( -\frac{\partial _{t}\Delta }{4kA}%
\left( -\frac{\partial _{t}A}{A}+2\frac{\partial _{t}\left( \frac{\partial
_{t}\Delta }{\Delta }\right) }{\left( \frac{\partial _{t}\Delta }{\Delta }%
\right) }\right) \right) =1
\end{equation}
Making it so, we obtain the RC vector in the same form that appears in the
paper \cite{NPV}, and also the \textit{integrability condition} (eq.2.13 of
that paper).

\subsubsection{\textbf{Second Example:}}

In the paper of R. Chan et al. \cite{Chan}, for the case of null flux of
heat, the following line element is considered:
\[
ds^{2}=-dt^{2}+B^{2}(t)[dr^{2}+r^{2}(d\theta ^{2}+sin^{2}\theta \ d\phi
^{2})]
\]
where:\ $B(t)=\frac{M}{2b}u^{2}$\ and\ $u=(\frac{6t}{M})^{1/3}$.

This is a FRW metric with $k=0$. The Ricci Tensor is diagonal and its
components are:
\begin{equation}
R_{tt}=\frac{2}{3t^{2}}\ ,\hspace{0.1in}R_{rr}=\frac{(6M^{2})^{1/3}}{%
b^{2}t^{2/3}},\hspace{0.1in}R_{\theta \theta }=R_{rr}r^{2},\mathrm{and\quad }%
\hspace{0.1in}R_{\phi \phi }=R_{\theta \theta }sin^{2}\theta .
\end{equation}
In this case: $\partial _{r}A=0$ and $\partial _{r}B=0$. It belongs to the
family number 7 of TABLE 1. Making use of equations (\ref{ColvectF7}), we
obtain the following expressions for the components of the RC: 
\begin{equation}
_{(7)}\xi ^{t}=\sqrt{\frac{3}{2}}\,t\hspace{0.2in}\mathrm{and}\hspace{0.2in}%
_{(7)}\xi ^{r}=\frac{r}{\sqrt{6}}
\end{equation}
This vector generates a Homothetic motion: $\pounds _{\xi }g_{ab}=\alpha
g_{ab}$ with $\alpha =\sqrt{\frac{3}{2}}$

\subsubsection{\textbf{Third Example:}}

Consider the metric \cite{NPV}:\newline
\[
ds^{2}=dt^{2}-R^{2}(t)(dr^{2}+r^{2}d\theta ^{2}+r^{2}sin^{2}\theta d\phi
^{2})
\]
where: $R(t)=\beta t^{\alpha }$,$\alpha \neq 1$ and $\beta $ and $\alpha $
are constants.

This is also an example of a FRW metric whose Ricci tensor components are:
\begin{eqnarray}
R_{tt} &=&A=-3\alpha (\alpha -1)t^{-2},\hspace{0.2in}R_{rr}=B=\beta
^{2}t^{2\alpha -2}(3\alpha ^{2}-\alpha ), \\
&&  \nonumber \\
R_{\theta \theta } &=&C=\beta ^{2}r^{2}t^{2\alpha -2}[\alpha (\alpha
-1)+2\alpha ^{2}],\hspace{0.2in}R_{\phi \phi }=Csin^{2}\theta .
\end{eqnarray}

Again, in this case $\partial _{r}A=0$ and $\partial _{r}B=0$, and it
belongs to Family 7 of TABLE 1. From equations (\ref{ColvectF7}) we get the
components of the RC vector:
\begin{equation}
_{(7)}\xi ^{t}=c_{1}\,t,\hspace{0.2in}\mathrm{and\quad }_{(7)}\xi
^{r}=-c_{2}\,r
\end{equation}
where 
\begin{equation}
c_{1}=\frac{1}{\sqrt{-3\alpha (\alpha -1)}}\qquad \mathrm{and\quad }c_{2}=-%
\frac{(\alpha -1)}{\sqrt{-3\alpha (\alpha -1)}}
\end{equation}
are constants.

This vector generates a Homothetic motion: $\pounds _{\xi
}g_{ab}=g_{ab}\delta $, with $\delta =\frac{1}{\sqrt{-3\alpha (\alpha -1)}}$

\section{\textbf{Final Comments.}}

In this paper is presented the form of the Ricci Collineation Vector in the
case of a diagonal, non-degenerate and spherically symmetric Ricci Tensor
for space-times that would admit a proper RC. It also has been shown the way
how to make the calculation of the RC for each one of the 64 families of
Ricci tensors obtained. The way is open for the exploration of the
non-diagonal and the degenerated Ricci tensors.

\section{\textbf{Acknowledgment.}}

This work has been partially supported by Consejo Nacional de
Investigaciones Cient\'{\i}ficas y Tecnol\'ogicas (CONICIT),Venezuela, under
contract number 4179476 T-217-93.\newline
The authors would also like to thank Dr. J. Carot for valuable suggestions.


\begin{thebibliography}{99}
\bibitem{Katzin69}  G.H. Katzin, J. Levine. \textit{J. Math. Phys}. \textbf{%
10},(4),(1969).

\bibitem{Paqui93}  A.H. Bokhari, A. Qadir. \textit{J. Math. Phys}. \textbf{34%
},(8),(1993).

\bibitem{Paqui94}  M.J. Amir, A.H. Bokhari, A. Qadir. \textit{J. Math. Phys}.%
\textbf{35},(6),(1994).

\bibitem{Paqui95}  T. Bin Farid, A. Qadir and M. Ziad. \textit{J. Math. Phys}%
.\textbf{36},(10),(1995).

\bibitem{Paqui96}  A.H. Bokhari and A.R. Kashif. \textit{J. Math. Phys}.%
\textbf{37},(7),(1996).

\bibitem{Bertol}  R. Bertolotti, G. Contreras, L.A. N\'{u}\~{n}ez, U.
Percoco, J. Carot. \textit{J. Math. Phys}. \textbf{37},(2), (1996).

\bibitem{Carot97}  J. Carot, L.A. N\'{u}\~{n}ez and U. Percoco. \textit{Gen.
Rel and Grav}.\textbf{29},10,(1997).

\bibitem{Kramer}  D. Kramer, H. Stephani, M. MacCallum. \textit{Exact
Solutions of Einstein's Field Equations.} Cambridge University Press(1980)

\bibitem{NPV}  L.A. N\'{u}\~{n}ez, U. Percoco and V. Villalba. \textit{J.
Math. Phys}.\textbf{31},1,(1990).

\bibitem{Chan}  R. Chan, J.P.S. Lemos, N.O. Santos, J.A. de F. Pacheco. 
\textit{Astroph. J}.,\textbf{342}, 976,(1989)
\end{thebibliography}
\end{document}